MRF-ZOOM: A Fast Dictionary Searching Algorithm for Magnetic Resonance Fingerprinting

Ze Wang

[1] Center for Cognition and Brain Disorders and the Affiliated Hospital, Hangzhou Normal University, Hangzhou, 310015, China

[2]Departments of Psychiatry and Radiology, Perelman School of Medicine, University of Pennsylvania

*Correspondence: Ze Wang, Center for Cognition and Brain Disorders and the Affiliated Hospital, Hangzhou Normal University, Hangzhou, 310015, China, Email: redhatw@gmail.com

*Abstract*

Magnetic resonance fingerprinting (MRF) is a new technique for simultaneously quantifying multiple MR parameters using one temporally resolved MR scan. But its brute-force dictionary generating and searching (DGS) process causes a huge disk space demand and computational burden, prohibiting it from a practical multiple slice high-definition imaging. The purpose of this paper was to provide a fast and space efficient DGS algorithm for MRF. Based on an empirical analysis of properties of the distance function of the acquired MRF signal and the pre-defined MRF dictionary entries, we proposed a parameter separable MRF DGS method, which breaks the multiplicative computation complexity into an additive one and enabling a resolution scalable multi-resolution DGS process, which was dubbed as MRF ZOOM. The evaluation results showed that MRF ZOOM was hundreds or thousands of times faster than the original brute-force DGS method. The acceleration was even higher when considering the time difference for generating the dictionary. Using a high precision quantification, MRF can find the right parameter values for a 64x64 imaging slice in 117 secs. Our data also showed that spatial constraints can be used to further speed up MRF ZOOM.

*Introduction*

Magnetic resonance fingerprinting (MRF) is a new technique for simultaneously quantifying multiple MR parameters using one temporally resolved MR scan [1, 2]. In MRF, random excitation flip angles, echo times (TE), and repetition times (TR) are used to make the temporal MR signal evolutions controlled by one parameter incoherent to those by others, so that different MR parameters will produce different MR timecourses. Those unique "MR fingerprints" are then matched to their most similar dictionary partners in a predefined MR signal database. The MR parameters generating the best match will be taken as the quantification results. This process does not involve any inversion problem as in the conventional data-fitting based quantification methods, so it avoids the inversion-related noise amplification. Major benefits of MRF also include: it is scale-free, capable of quantifying several parameters simultaneously, and not sensitive to within-plane motions [1]. While all those features are appealing to clinical applications, a practical issue is that we need a memory and computationally efficient dictionary generating and searching (DGS) algorithm. The original MRF DGS is an exhaustive brute-force approach [3], which is both storage space demanding and time consuming. Its multiplicative computation complexity (MCC) can easily make it intractable for any state-of-art computers when more parameters or higher precision or larger value scan range are designated.

The purpose of this paper was to develop a fast and efficient MRF DGS algorithm. Given the MR fingerprints, the MCC of DGS is contributed by several factors: the number of parameters n, the quantification precision p, the data acquisition length t. Similar to a multi-dimensional Fourier transform, multiple-parameter induced MCC can be reduced to an additive computation complexity (ACC) if MR parameters can be processed independently. The resolution-related MCC can be dramatically reduced if a convexity-based searching process can be used. To exploit these possibilities, we empirically characterized the properties of the dictionary matching error function by simulations and based on those properties we proposed an MR parameter separable (PS) multi-resolution (MS) MRF DGS algorithm. For the simplicity of description, this algorithm was dubbed as PS-MRF-ZOOM or MRF ZOOM henceforth.

*Theory*

*MR signal time evolutions and MRF dictionary generating and searching (DGS)*

To get a unique MR "fingerprint", the temporal evolutions of MR signal need to be different for different material (or tissue) MR parameter values [1]. In MRF, this is made possible by randomly changing the acquisition parameters at random time in order to get nearly random but incoherent temporal evolution patterns of each underlying MR parameter. Spatial incoherence is also introduced to further enhance the uniqueness of the acquired MR signal at each particular voxel. After acquiring MR fingerprints, MRF generates a dictionary of ideal MR fingerprints based on Bloch equations [4] by varying the modeled MR parameters within a certain range of values but using the same acquisition parameters such as RF flip angles, TEs, TRs, as those used during data acquisition. Denoting the acquired MR fingerprint and the dictionary data by $\vec{M}_a = (M_{ax}, M_{ay})$ and $\vec{M}_d = (M_{dx}, M_{dy})$, respectively (the subscripts x and y mean the two orthogonal component of the transaxial magnetic signal along the x and y axis, respectively), the goal of MRF dictionary searching is to minimize the distance function $F(\vec{M}_a, \vec{M}_d)$, which can be either the Euclidean distance or Pearson correlation coefficient (CC). Similar to [1], we used CC in this paper, and the optimization problem of MRF dictionary searching is then to maximize the CC between the MR fingerprints and the dictionary entries. For two complex timecourses $\vec{M}_a = A_0 e^{i\theta_0}, A_1 e^{i\theta_1}, ...$ and $\vec{M}_d = B_0 e^{i\varphi_0}, B_1 e^{i\varphi_1}, ...$ (where $A_j$ and $\theta_j$ means the magnitude and phase of $\vec{M}_a$; $B_j$ and $\varphi_j$ means the magnitude and phase of $\vec{M}_d$. j means j-th timepoint. j=0, … N, N is the data length), their CC (the absolute value here) can be defined by [5]:

$$CC(\vec{M}_a, \vec{M}_d) = \left| \frac{\sum_{j=0}^{N} A_j B_j e^{i(\theta_j - \varphi_j)}}{\sqrt{\sum_j^N A_j^2} \sqrt{\sum_j^N B_j^2}} \right| \tag{1}$$

Because the same imaging parameters are used for acquiring $\vec{M}_a$ and generating $\vec{M}_d$, $CC(\vec{M}_a, \vec{M}_d)$ can be simplified as a function of MR parameters to be determined. In this paper, only T1, T2, df (off-resonance frequency) and the proton density are considered. As proton density can be quickly determined from the signal scale, CC reduces to a function of T1, T2, and df. Method extensions for more parameters should be straightforward.

To reduce computation time, both MR fingerprints and dictionary entries can be normalized to have a unit norm [1], so that we can skip calculating the denominator of Eq. 1. Therefore, $CC(\vec{M}_a, \vec{M}_d)$ of an arbitrary $\vec{M}_d$ can be re-described as $CC(T_{1d}, T_{2d}, df_d)$, and further simplified as $CC(T_1, T_2, df)$ without introducing ambiguity. For the convenience of method derivation, $CC(T_{1d}, T_{2d}, df_d)$ can be re-expressed as a function of the parameter difference: $f_{cc}(\Delta T_1, \Delta T_2, \Delta df)$, where Δ means the difference between the dictionary parameter and that of the MR fingerprints. Because both the fingerprints and dictionary entries are unique to specific values of T1/T2/df, $CC(T_1, T_2, df)$ will have only one global optimum in the space spanned by $(T_1, T_2, df)$. In other words, $f_{cc}(0,0,0) > f_{cc}(\Delta T_1, \Delta T_2, \Delta df)$ for any $\Delta T_1 \neq 0, \Delta T_2 \neq 0, \Delta df \neq 0$.

Assuming that the fingerprint length is T and maximal possible number of steps for the 3 parameters are R, S, Q, respectively, the original DGS has a computation complexity of $O(RSQ9^T)$, where R, S, and Q depends on the searching range and resolution, 9 comes from the 9 times multiplications and 9 additions in Bloch equation (see Appendix). The complexity of CC calculation due to the RSQT times multiplications and additions $O(RSQT) < O(RSQ9^T)$ and then is ignored when the overall complexity is considered. When R, S, Q, and T approaches infinity (for the convenience of estimating the computation complexity), they can be simply replaced by one input size variable n and the complexity can be simplified as $O(n^3 9^n)$.

*Empirical properties of $CC(T_1, T_2, df)$*

The key to reduce the computation burden of MRF DGS is to downgrade the multiplicative computation complexity into an additive one. According to the concept of partially separable functions, $CC(T_1, T_2, df)$ can be approximately represented by a multiplication of a function of $T_1$, a function of $T_2$, and a function of $df$. If the individual functions are mutual independent, the maximum of CC can be then found by 3 mutual independent processes, each for one component function. However, since the MRF signal doesn't have an explicit analytic formula, it is difficult to derive such a partial separation formation for the CC function. We rather explicitly assessed the partially separable property of CC by simulations, and then derived a pseudo separable DGS algorithm below.

*Property I.* For the simplicity of descriptions, let's assume a uniform norm for both the fingerprints and dictionary entries and $df_a = 0$ (which can be always met by rotating the coordinate system using the Larmor frequency calculated from the local field strength (after including field inhomogeneity)). Therefore, CC defined in Eq. 1 is the same as an inner product of the two complex vectors which is $\leq 1$ (equals 1 if and only if the two vectors are the same). For any $\Delta T_1$ and $\Delta T_2$, if there is no off-resonance difference ($\Delta df = 0$), Eq. 1 will reduce to $CC(\vec{M}_a, \vec{M}_d) = \sum_{j=0}^{N} A_j B_j$, which will be greater than 0. But if there is a non-zero $\Delta df$, the off-resonance difference will induce phase difference between the fingerprint and the dictionary entry, which will be further incoherently changing during its time evolution in the random MRF TRs. This random and incoherent phase perturbation will occur to all timepoints included in the summation operation of Eq. 1, and will inevitably cause a phase cancellation for the summation process and subsequently reduce CC. This phenomenon (*Property I*) can be summarized by: $f_{cc}(\Delta T_1, \Delta T_2, \Delta df) < f_{cc}(\Delta T_1, \Delta T_2, 0)$, which says that CC reaches the global maximum only at $\Delta df = 0$ independent of $\Delta T_1$ and $\Delta T_2$.

*Property II.* Since the off-resonance induced phase is pseudo-periodic with respect to TR, the phase cancellation effects can present periodic patterns with respect to off-resonance though CC should still decay with increasing $|\Delta df|$ as described in Property I. With no TR variations, the period of $f_{cc}(\Delta df)$ will be $\omega = 1/TR$ because $\Delta df = \omega$ will contribute a $2\pi$ rotation which will not change the original signal. Meanwhile, the random phase variations due to the random variations of TRs will result in a gradually decaying CC when $|\Delta df|$ increases.

*Property III.* The above mentioned CC properties suggest that $df$ can be identified independent of other parameters. Once $df$ is known or even close to be the actual value, CC or $f_{cc}(\Delta T_1, \Delta T_2, \Delta df)$ will likely be a smoothly decaying function when $|\Delta T_1|$ and $|\Delta T_2|$ increase. Because of the random RF flip angles and random TRs used in data acquisition, different T1 or T2 caused temporal signal evolutions are incoherent. Also because larger $|\Delta T_1|$ and $|\Delta T_2|$ will cause larger signal intensity differences and subsequently make the dictionary MR timecourse more different from the fingerprints, it is reasonable to posit a convexity property of CC vs $|\Delta T_1|$ and $|\Delta T_2|$ (*Property III*). In addition, CC vs $|\Delta T_1|$ and $|\Delta T_2|$ should approximately present a unimodality property in order to have a unique MRF match. Based on the convexity of the CC function, we can get that for any $\Delta T_2$, $f_{cc}(\Delta T_1, \Delta T_2, \Delta df)$ will peaked at only one position close to $(0, \Delta T_2, \Delta df)$. A similar feature can be postulated for T2.

*MRF ZOOM: a parameter separable (PS) multi-resolution (MS) MRF dictionary generation and matching algorithm*

Property I suggests that $df$ can be fingerprinted separately from T1 and T2. Once $df$ is determined, T1 and T2 can be determined independently in two parallel threads. This PS process will break the multiplicative MRF computation complexity into an additive one.

$df$ can be identified using the following searching steps: 1) Find the initial CC maximum and the associated $df$ value using a brute-forcing scan (checking all possible values) with a searching step smaller than the pseudo-period $\omega$; 2) reduce the searching range to be $2\omega$ and centered at the initial optimum $df$ and repeat step 1 with a smaller step like $\omega/20$; 3) starting from the optimum $df$ identified in 2), repeat step 1 using a searching interval of $\omega$ until reaching the boundaries of the original searching range; 4) reduce the searching range to be $-0.5\omega \sim 0.5\omega$ and centered at the previous optimum $df$ and repeat step 1 with the finest searching step (resolution).

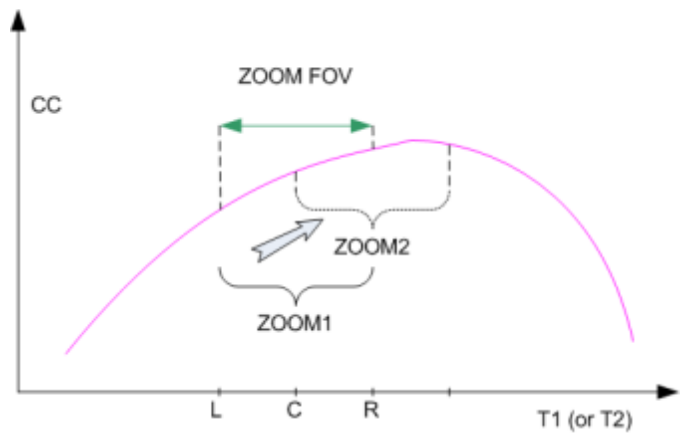

Fig. 1. Illustration of a 1D MRF ZOOM. Starting from an initial place as marked by C (ZOOM center), MRF ZOOM compares the left (L) and right (R) end of the current ZOOM FOV with the center and move the ZOOM to be centered at the maximum (in terms of maximal CC) of the 3 locations until C remains the maximum. In this figure, MRF ZOOM starts from the location of ZOOM1 and moves to ZOOM2, similar to a hill-climbing process as indicated by the arrow.

The 3rd CC property (convexity and unimodality with respect to T1 and T2) suggests a zoom-like multi-resolution searchlight algorithm for identifying T1 and T2 either separately or together. This algorithm is similar to the well-known golden section searching algorithm [6]. To reduce computation time for generating the dictionary items and comparing them with MR fingerprints, we need to reduce the number of comparison steps. Given a pre-specified range for the parameter (T1 or T2), the MRF ZOOM can be started from the center of the searching range with a coarse resolution to quickly locate the optimal section where the maximum CC exists. Fig. 1 illustrates an MRF ZOOM scheme for one parameter with one resolution (the ZOOM field of view (FOV)). Starting from an initial ZOOM center (C) that can be the mean of the entire searching range or the parameter value from the neighboring voxel, MRF ZOOM first finds the maximum CC from the three positions: the ZOOM center (C), the left (L) and the right (R) end points. It then moves the ZOOM to be centered at the new maximum point and repeat the same ZOOM moving process until that CC at the center is greater than CC's of the two ends. For the purpose of illustration, MRF ZOOM is moved only once in Fig. 1, but it can be moved several times in real situations. To quickly locate the optimum section, a big initial ZOOM can be used. For example, if the initial ZOOM FOV is ½ of the entire searching range, starting from the center of the entire range, there will be at most 2 requests for generating 2 dictionary entries and CC calculations and 3 CC comparisons. Note that we always assume that CC of the current ZOOM center is known. Once the current ZOOM stops moving, its FOV can be reduced to enter a finer resolution zooming process until a pre-specified resolution is met.

In case of a variation to the unimodality property (property III) of CC with respect to T1 and T2 ( $f_{cc}(\Delta T_1, \Delta T_2)$ with a given $\Delta T_2$ may not peak at $(0, \Delta T_2)$ and $f_{cc}(\Delta T_1, \Delta T_2)$ with a given $\Delta T_1$ may not peak at $(\Delta T_1, 0)$), independent MRF ZOOMs for T1 and T2 may not end at the global optimum (0,0). One solution for this problem is to use several interleaved T1 and T2 ZOOMs so the previously identified T1 or T2 can be used to bring the next T2 or T2 ZOOMs closer to the global peak. The other is to use a 2-dimensional (2D) ZOOM to find the optimal T1 and T2 at the same time. Fig. 2 illustrates a zooming process of a 2D MRF ZOOM with a fixed resolution pair (one for T1 and the other for T2). Similar to the 1D ZOOM, the ZOOM center is compared with the left, right, up, and down boundary points (marked by the green circles) as well as the 4 corner points (marked by the purple squares). The one with the maximal CC will be set as the new center of the 2D ZOOM, and the same comparison process will be repeated until the 2D ZOOM covers the optimal 2D area where the optimal T1 and T2 reside in. The ZOOM FOV can be then gradually reduced to find the final solution with the specified precision (resolution) as schemed by the black boxes in Fig 2.

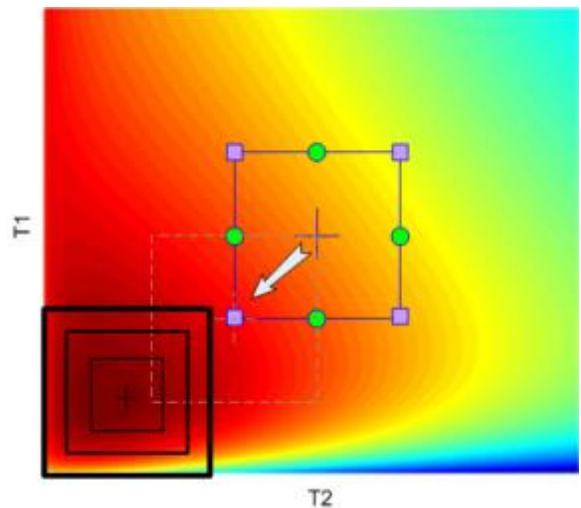

Fig. 2. Illustration of a 2D MRF ZOOM. With an initial searching resolution, MRF ZOOM starts dictionary generation and matching in the center of the original searching range. It will then move the ZOOM to be centered at the maximum of the boundary points (the green circles and purple squares) and repeat the above matching process until reaching a place (the thick black big square in the figure) where the ZOOM center has a CC greater than the 4 boundary positions (marked by green circles). Then a finer resolution is used to repeat a new MRF ZOOM

process until the specified resolution is reached. The white arrow indicates the ZOOM moving direction; the green circles indicate the boundary points that are closest to the center; purple squares indicate the corner points. The dashed big square indicates an intermediate location of ZOOM.

For both 1D ZOOM and 2D ZOOM, the unimodality and convexity of $f_{cc}(\Delta T_1, \Delta T_2)$ can be used to reduce the number of comparisons and the associated dictionary entry generation and CC calculations. For example, in 1D ZOOM, if the right end is greater than the center (as shown in Fig. 1), we can directly move the ZOOM to the right end and don't need to visit the left end at all because the left end should have smaller CC than the center (C) and the right (R) end. For 2D ZOOM, a similar strategy can be used. For example, 2D ZOOM can evaluate two boundary points along two dimensions first. Let's say L and the down (D) side neighbors. If both of them are greater than the center (in terms of CC) like the case in Fig. 2, the 2D ZOOM can be directly moved to the down-left corner and the number of dictionary entry generations and CC calculations will be 3 rather than 9. If both L and D are smaller than C, the other 2 neighbors: R and the up (U) side one, will be evaluated. If both R and U are smaller than C, current ZOOM moving will be stopped. And the total number of dictionary entry generations and CC calculations will be 4. If both R and U are greater than C, ZOOM will be moved to the up-right corner and the neighbor comparisons will continue. But for current ZOOM moving, the number of dictionary entry generations and CC calculations will be 5. If one of the first 2 neighbors is greater than C but the other is smaller, ZOOM will be moved to a neighbor next to the one greater than C. Assuming L>C and D<C, we will compare U and C. If U>C, ZOOM will be moved to the up-left corner or else it will be moved to L. The former case needs 4 times dictionary generation and CC calculation; the latter only needs 3, which will be the same for the other case of L<C and D>C.

For df searching alone, the computing complexity is $O(n9^n)$ without prior dictionary and $O(n)$ with prior dictionary. The complexity of T1/T2 ZOOM part is smaller than the complexity of two independent searching process for T1 and T2 respectively, which is $O(\log(n)9^n)$. The overall complexity can be then approximated by $O(n9^n) + O(\log(n)9^n) \approx O(n9^n)$ without using prior dictionary for df and by $O(n) + O(\log(n)9^n) \approx O(\log(n)9^n)$ if prior dictionary is used for df.

In summary, the proposed PS-MRF-ZOOM uses a resolution adjustable ZOOM to find df first and then T1 and T2. The entire algorithm complexity is much smaller than the original exhausting searching algorithm. Similar analysis can be performed to develop a corresponding algorithm if there are more than 3 parameters to be identified.

*Methods*

For the simplicity of method descriptions, the inversion-recovery balanced steady state free-precession (IR-bSSFP) sequence as schemed in [1] was assumed in this paper. Nevertheless, extension of MRF ZOOM for other sequence-based data acquisitions should be straightforward.

All following computation algorithms and processes were implemented in C++, and the experiments were performed in a laptop with 8 GB memory and a 2.0 GHz dual-core CPU.

*Experiment 1: CC mapping using the brute-force MRF DGS algorithm*

Synthetic MR fingerprints were created using Bloch equations (Eq. 1 and 2 in Appendix) with 1400msec/500msec/100Hz. The IR-bSSFP sequence was simulated using the same settings as in [1]. RF flip angles between 0~79° were generated using Perlin noise [7]; RF phases were oscillating between 0°, 90°, and 180°. TRs were determined from Gaussian noise after being remapped to be from 0 msec to 6 msec plus a minimum TR (14 msec) for data acquisition. The original exhaustive DGS method was used to generate the dictionary. To limit the time for computation, the parameter ranges were set to be 100~5500 msec, 50~1200 msec, -300~300 Hz for T1, T2, and df, respectively. Dictionary entry length was 500 timepoints. Dictionary resolutions for T1, T2, and df were 10 msec, 10 msec, and 1 Hz, respectively. CCs between the synthetic fingerprints and the 37260000 dictionary entries were calculated.

*Experiment 2: MRF ZOOM evaluation 1*

MRF ZOOM was compared with the original brute-force algorithm for identifying 250 synthetic MR fingerprints generated with 250 different T1/T2/df values selected from the range of 500~2000 msec/200~800 msec/-30~450Hz, respectively. The same resolutions as in Experiment 1 were used in the brute-force searching. The searching ranges for T1/T2/df were reduced to be 500~2000 msec/200~800 msec/-30~450Hz, respectively, resulting in a 16.1 GB dictionary plausible to be saved in the computer's hard disk. MRF ZOOM was performed using the same resolution with or without a pre-generated dictionary. Computation time was recorded as a performance index. Number of dictionary matching was also recorded in MRF ZOOM.

Detailed zooming procedures were:

1) df was searched using the algorithm as described in Theory. $\omega$ was set to be 70 Hz (~1000/14, 14 msec is the minimum TR used in this paper), and the initial T1/T2 were set to be 1000msec/500msec. a) The entire searching range was scanned with a step of 60Hz to find the initial CC maximum and the associated $df$ value. b) The searching range was reduced to 150Hz around the initial optimum $df$, and another exhaustive search with a step of 3Hz was used to find the new CC maximum and the optimum $df$. c) Locations in the original searching range away from the new optimum with any integer times of $\omega$ were evaluated to update the maximum. Zooming stopped if either of the following criteria was met: (1) finished searching with dfres=1 Hz (the finest resolution used); (2) the maximum CCs of 3 different resolution-based zooming processes were nearly the same up to a difference < 1e-7.

2) A 2D MRF ZOOM with an initial resolution of 200msec and then 100 msec for T1/T2 was used to find a tentative T1/T2. The tentative df was used for all dictionary matching.

3) A 1Hz stepwise brute-force search was performed within a range of $-0.5\omega \sim 0.5\omega$ around the tentative optimal df identified in 1) to update df. The optimal T1/T2 identified in 2 were used.

4) T1 zooming was performed using the optimal df identified in 3) and the optimal T2 found in 2) with an initial resolution of 100 msec. The temporary optimal T1 and df were then used for T2 zooming with an initial resolution of 100 msec. Three sequential T1 zoomings were performed with the updated optimal T2 with a resolution of 50, 20, and 10 msec, respectively. Similar processes were performed for T2 using the new optimal T1.

5) Final T1/T2 zooming with a resolution of 1 msec/1 msec was performed using the 2D MRF ZOOM.

The same procedures were repeated with different initial values of T1/T2: 500msec/200msec, 800msec/500msec, 2000msec/800msec, to evaluate the sensitivity of the algorithm to the initial values. $\omega$ was also changed from the minimal resolution (1Hz here) to 69Hz to assess the sensitivity of the df searching process. When $\omega$=minimal resolution (1Hz), the df searching process was the same as a brute force searching.

*Experiment 3: MRF ZOOM for a 64x64 brain image slice*

Synthetic T1 and T2 parameter maps (64x64 voxels) were generated based on a high resolution T1-weighted structural image. T1/T2/off-resonance ranges were set to be 800~3201

msec/50~601 msec/-48~85Hz, respectively. Precisions for T1/T2/df were set to be 1 msec/1 msec/1 Hz, respectively. MR fingerprints were generated at each of the 1731 intracranial voxels using Bloch equations. The same MRF ZOOM algorithm as described above was used to determine T1, T2, and df for each voxel.  To further improve parameter determination speed, a second MRF experiment was performed by including parameters already identified from neighboring voxels as priors.  Suppose a preceding neighboring voxel has a T1/T2/df of T1p/T2p/dfp. Ranges of T1p±3000, T2p±800, dfp±150 were generated and their overlap with the original T1/T2/df searching ranges were used as the final ranges for MRF ZOOM for the current voxel.

*Experiment 4: noise effects on MRF*

Noise contamination is inevitable in MRI. To assess the effects of noise on MRF, different random white noise was added to the real part and imaginary part of the same MR fingerprints as in Experiment I. 51 different noise levels were used to yield a range of cross-correlation (0.04-0.98) between the noise-free MR fingerprints and the noise contaminated fingerprints. Both the brute-force DGS algorithm and MRF-ZOOM were used to find the designated 3 parameters: T1/T2/df. The same parameter searching range as in Experiment 1 was used for all MRF-ZOOM experiments. To save computation time for the brute-force DGS processes, only the first 3 DGS processes where the noise levels were highest and the one with a modest CC (0.4)between the MR signals with and without noise  were performed using the same searching range as in Experiment 1. After that, MRF-ZOOM was applied first to find a set of solution: T1zoom/T2zoom/dfzoom and then a smaller searching range (T1zoom±550, T2zoom±300, dfp±50) was used for the brute-force DGS. MRF quantification errors were measured by the difference between the identified value and the ideal value.

Temporal smoothing was applied to both the noise contaminated MR fingerprints and MR dictionary and the above processes were repeated to evaluate whether noise suppression can improve MRF parameter quantification accuracy or not. 3 points and 5 points moving Gaussian smoothing were tested.

**Results**

*Experiment 1: CC mapping.* For the parameter ranges of 100~5500 msec, 50~1200 msec, -300~300 Hz and a resolution of 10 msec/10 msec/1 Hz, for T1/T2/df, respectively, the brute-force MRF algorithm took 4 hours and 35 minutes to generate the dictionary and calculating the CC maps. Fig. 3A shows the CC versus (vs) df curves of all evaluated pairs of T1/T2 values. It clearly evidenced Property I of CC map as postulated in Theory: CC(df) reaches the global peak at the designated off-resonance value, which is the only one global optimum. Property II of CC vs df was also clearly demonstrated in the figure. Fig. 3B shows the CC(T1, T2) map using $\Delta df$ =0. The global peaks of CC(T1) for any T2 located nearly the same T1 location. Similar observations can be found for CC(T2). These patterns support the validity of the algorithms derived from the aforementioned CC properties.

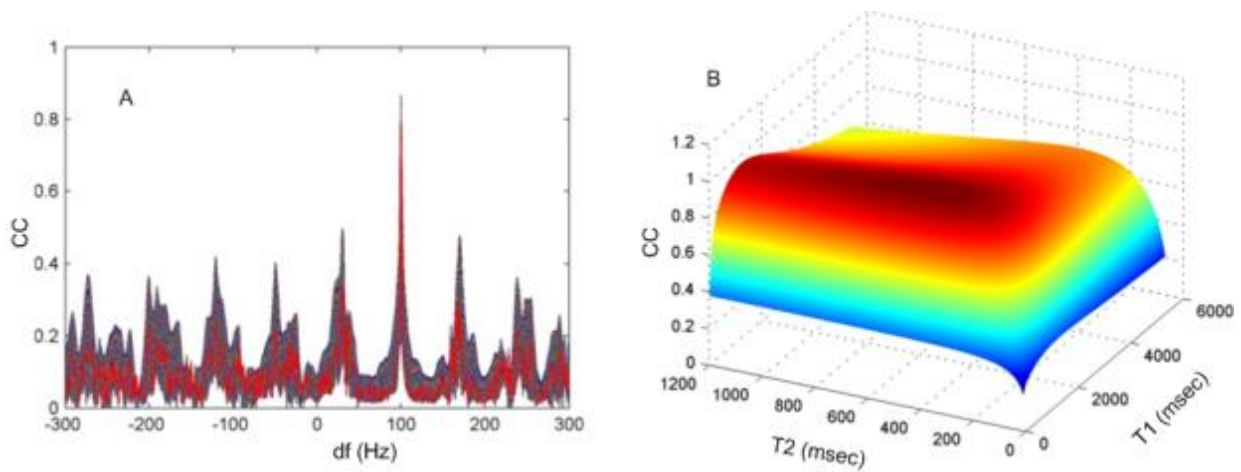

Fig. 3. CC as a function of df, T1, and T2. The MR fingerprints were generated using T1/T2/df=1400/500/100 (msec/msec/df). A) CC vs df curves of all evaluated T1/T2 values, B) CC vs (T1, T2) map at df=100 Hz.

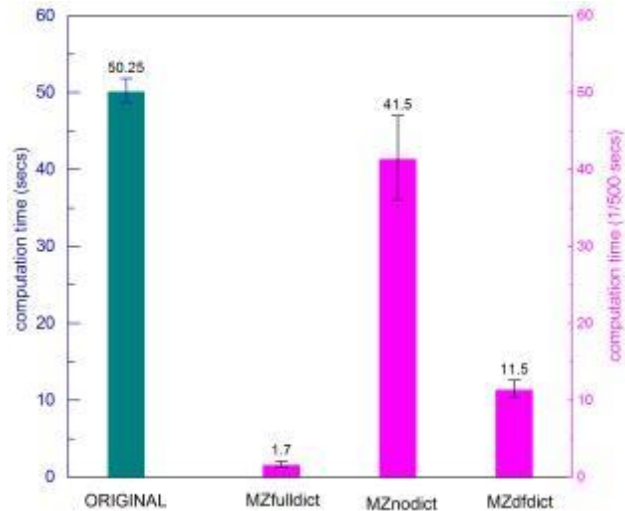

Fig. 4. Computation time of different MRF parameter quantification algorithms. ORIGINAL means the original brute-force algorithm, MZfulldict means MRF ZOOM with a full pre-defined dictionary, MZnodict means MRF ZOOM without using any predefined dictionary, MZdfdict means MRF ZOOM with a predefined dictionary generated with a fixed T1/T2 but with different df values. Dictionary generation time for the prior dictionary (ORIGINAL, MZfulldict, MZdfdict) was not counted in the graph. Error bars mean standard deviation.

*Experiment 2: MRF ZOOM evaluation 1.*

Using the same resolution (10msec/10msec/1Hz for T1/T2/df, respectively) and the same searching ranges (500~2000 msec/200~800 msec/-30~450Hz for T1/T2/df, respectively), both the original brute-force searching algorithm and MRF ZOOM successfully identified the target parameter values, which were set to be an integer number of the resolution. Fig. 4 shows the average time (including the searching time and disk access time) used for T1/T2/df determinations for the 250 fingerprints. The original brute-force method took 50.25±1.54 (mean ± standard deviation) secs to find one set of parameters, not including the 1980.84 secs for generating the dictionary. MRF ZOOM took 0.0034±0.0007 secs (14779 times faster) to determine all the parameters if the pre-defined dictionary was used, 0.083±0.011 secs (605 times faster) if no prior dictionary was used. 189±25 dictionary entries were checked before MRF ZOOM found the final solution. Using the pre-defined df dictionary, which took 0.29 secs to create, MRF ZOOM took 0.023±0.0022 secs (2184 times faster) for finding the parameter values. MRF ZOOM with different initial values successfully identified the right answer.

*Experiment 3: MRF ZOOM for a brain image slice*

Fig. 5 shows the results of MRF ZOOM for the synthetic fingerprints of one brain image slice. With the pre-defined dictionary with changing df but a fixed T1 and T2, MRF ZOOM took 116.72 secs to converge. Using the preceding nearest neighbor voxel as a constraint, MRF ZOOM used 89.81 sec to converge. Both processes yielded the same results. From Fig. 5, MRF ZOOM identified T2 and df for all voxels (Fig. 5B and 5C) without any errors. For T1 (Fig. 5A), 10 voxels showed minor errors (between -1 to 3).

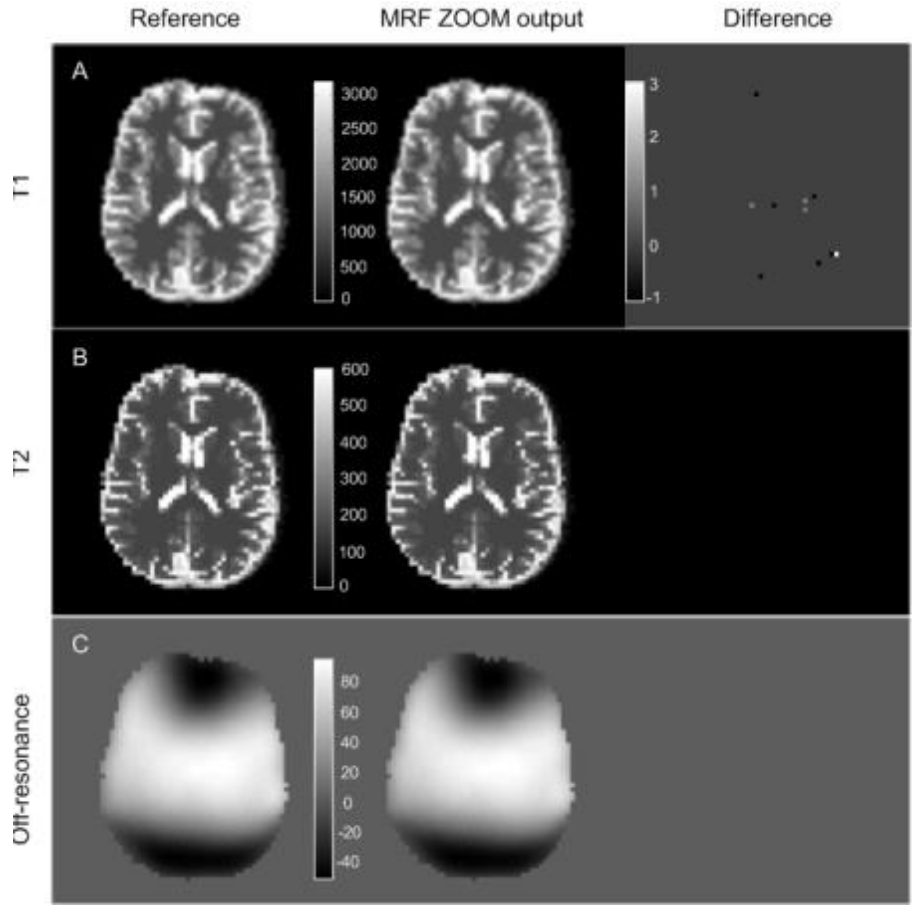

Fig. 5. MRF ZOOM results for one brain slice. The left column is the gold standard of T1/T2/df maps. The second column is the MRF ZOOM identified T1/T2/df maps. And the right column is the difference between MRF ZOOM results and the reference.

*Experiment 4: noise effects*

Fig. 6 shows the CC mapping results after adding a modest level of noise (resulting in a CC of 0.4 between the original signal and noise contaminated signal). Similar to the noise-free case (Fig. 3A), Fig. 6A demonstrates that even after adding modest noise, CC(df) has only one global optimum which is the same as the designated off-resonance value. Fig. 6B shows a similar convexity pattern to that in the noise-free case (Fig. 3B). Both T1 and T2 peaked at the only global optimum which is nearly the same as the designated locations. Fig. 7 shows the MRF quantification errors when different level of noise was added to the fingerprints. The brute-force DGS and MRF-ZOOM performed the same except the first testing point when heavy noise was added causing a CC of 0.038 between the noise-free and noise contaminated fingerprints (Fig. 7A). Further investigation revealed that the df searching step needed to be adjust by using finer initial searching step and larger relative searching range in the successive searching substeps in order to get the same df quantification results. And then the same T1 and T2 zooming process can be used to find the same T1 and T2 as in brute-force DGS. In Fig. 7B, a 3 point Gaussian smoothing applied to both the fignerprints and dictionary showed reduced quantification errors for nearly all noise level except for T2 at the first two highest noise levels. The improvement becomes smaller when the noise level goes lower.

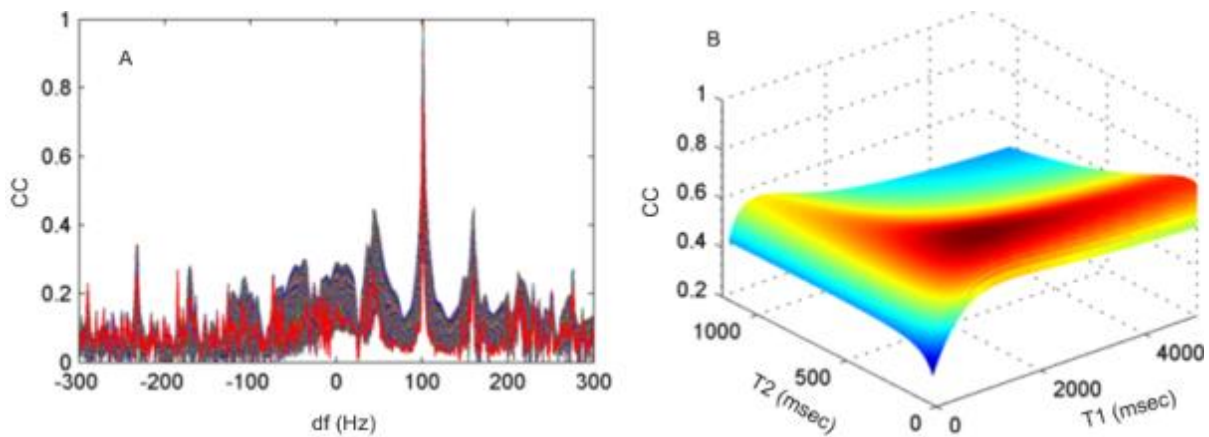

Fig. 6. CC as a function of df, T1, and T2 when noise presents in MR fingerprints. The MR fingerprints were generated using T1/T2/df=1400/500/100 (msec/msec/df). The noise was set to reduce the auto-CC (between the same MR fingerprints with and without noise) to be 0.4. A) CC vs df curves of all evaluated T1/T2 values, B) CC vs (T1, T2) map at df=100 Hz.

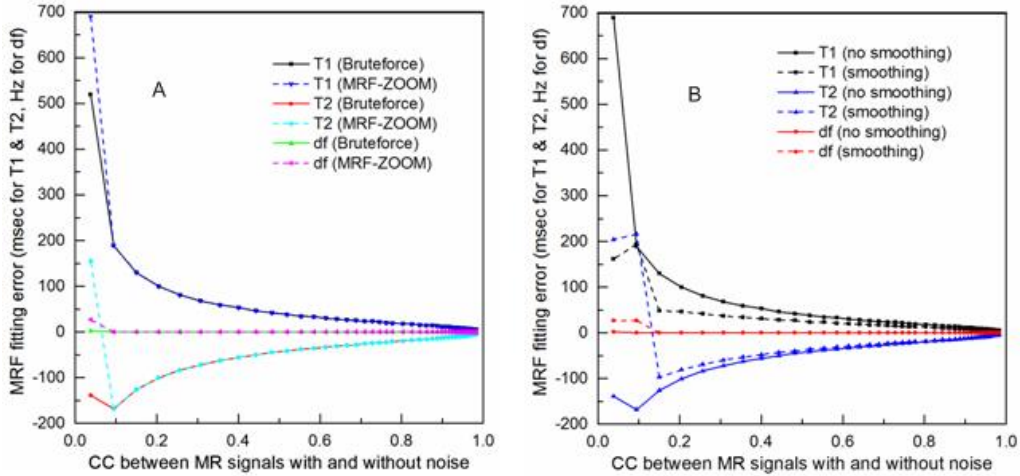

Fig. 7. MRF parameter quantification errors at different noise levels. A) quantification errors of T1/T2/df by different algorithms; B) quantification errors by MRF-ZOOM with and without 3 points Gaussian smoothing

**Discussion**

A super-fast MRF DGS algorithm: the MRF ZOOM was proposed in this paper. Based on empirical analyses of MRF data fitting objective function (here the Pearson correlation coefficient), MRF ZOOM uses two major steps to reduce computation complexity of MRF parameter quantification. First, it breaks the multiplicative computation complexity for quantifying multiple parameters into an additive one. In this paper, off-resonance was identified first due to its dominant effects on the matching distance function. An arbitrary T1/T2 of 1000/10 msec was used for this step. Different T1/T2's may yield different CC vs off-resonance curves. But they should still peak at the only one global optimum (Property I) as we can see from Fig. 3A. This property may not be true if T1 or T2 is too short as compared to the minimum TR (14 msec in this paper). In that case, a second searching process can be used after finding an approximate value for T1 and T2. The second step used in MRF ZOOM to reduce computation complexity is the convexity-based multi-resolution ZOOM. Both the empirical MRF matching distance function analysis and simulations showed that the distance function is unimodal and convex with respect to T1 and T2. This property suggests a fast-moving searchlight-like approach to quickly find the target parameter section with any given resolution, which is very similar to the way a camera was used to capture remote object, where a large ZOOM is first used to find the target range and then is gradually reduced to see details.

Using the same resolution, MRF ZOOM was fourteen thousands times faster than the original brute-force MRF algorithm when a pre-defined dictionary was available; 2184 times faster when the off-resonance dictionary was pre-defined; 605 times faster when no prior-dictionary was available. The dramatically reduced computation complexity of MRF ZOOM makes it possible to determine MR parameters with high resolution, which however is nearly impossible for the original searching algorithm due to the extremely large disk space demand and long computation time. When applied to synthetic MR fingerprints from a brain image slice with a high resolution and large searching range, MRF ZOOM identified T2 and off-resonance in 116.72 secs without any errors. It found the right T1 values for most voxels except 10 voxels with large T1 (>2400 msec). When checking the matching process, we found these minor errors were caused by the CC calculation errors for the normalized data. By using the Euclidean distance as the objective function or by using the non-normalized fingerprints and MR dictionary entries at the finest resolution, we were able to find T1 for all voxels without errors. In real situations, brain MR parameters present spatial correlations, which can be used to further increase MRF pattern matching speed. By using the adjacent voxel as constraint in MRF ZOOM, we showed a 20% computation time reduction.

Noise may affect MRF parameter quantifications and has not been assessed before. Based on additional analyses, we found that moderate to heavy additive random noise in the fingerprints reduced the CC magnitude but didn't change the aforementioned 3 properties, so both the brute-force DGS and MRF ZOOM can still find a unique global peak. However, noise contaminations caused errors to the identified T1 and T2, which increased with noise level. One possible approach to improve MRF quantification accuracy is denoising. Our data showed that temporal smoothing can reduce the quantification errors for most of the tested noise level. However, we also found that larger smoothing kernel (covering more than 5 successive timepoints) increased quantification errors instead (data not shown), indicating a greater suppression to the useful high-frequency components contained in MR fingerprints. If MRF is designed to have relatively less high-frequency components, it is likely that temporal smoothing would further improve MRF quantification accuracy and it would be less sensitive to the smoothing kernel length. In real application, noise can be measured independently and the same simulations as conducted in this paper can be performed to find the approximate range of the quantification accuracy, which can be used as a guidance to prospectively select the minimum resolution to be used for DGS. For example, if the noise contamination caused the auto-CC (between the same signal with and without noise) reduced to 0.5, the offset to T1/T2 would be around 40 msec, so an acceptable choice of the resolution for T1 and T2 can be 10 msec, which certainly will reduce DGS time as compared to the case of 1 msec.

It is worth to note that the 3 properties of CC with respect to df/T1/T2 were empirically derived from an analysis of the CC function though they were proven to be true in the CC maps. To test whether those properties are unique to the used random flip angles, RF phases, and TRs, we created another series of random flip angles, RF phases, and TRs, and repeated the CC mapping process and found similar patterns to those shown in the paper. It is possible that the 3 properties might not be completely true in certain cases using arbitrary MRF pulse sequences, but we can still use data simulations to find the patterns of the parameter quantification objective function (CC in this paper) and design a fast DGS algorithm accordingly.

One limitation of this paper was that we didn't have data acquired in real environment. In the original MRF work[1], the image was undersampled by tens of times, resulting in a very low signal to noise ratio. Impressively, the authors demonstrated that the brute-force DGS can still get the parameters accurately. One concern might be the noise-induced local maxima. But based on the linearity of dot product which is literally the same as the CC function used in this paper, adding the same noise or artifacts would add similar if not the same CC difference to all dictionary entries. In other words, noise or artifacts to fingerprints should not change the properties of the CC function because the dictionary doesn't change. This was shown in our noise-based simulations. Certainly, real data will be required to verify and improve the DGS algorithm in the future.

Although MRF ZOOM was only tested in 3 parameter determination MRF, similar concept can be extended for more parameter quantifications if both the fingerprints and the Bloch equations are modified appropriately. Pseudo-random sampling was used in [1] to enhance the uniqueness of the acquired MR fingerprints for each specific voxel location. This process is similar to adding spatially incoherent pseudo-random noise to the fingerprints. Adding random noise should be sufficient to assess its effects on MRF parameter quantification for each particular voxel.

In summary, MRF ZOOM solves a practical issue of MRF for multiple-parameter quantification with high resolution.

**Acknowledgement**

This work was supported by the Hangzhou Qianjiang Endowed Professor Program and the Youth 1000 Talent Program of China.

**Additional Information**

The author declares no competing interest regarding this paper.

## Appendix

In the rotating frame, the time evolution of the spin magnetization vector $\vec{M} = (M_x, M_y, M_z)^T$ can be modeled by Bloch equation [4]:

$$\frac{d\vec{M}}{dt} = \gamma(\vec{M} \times \vec{\Delta B}) - \begin{pmatrix} M_x/T_2 \\ M_y/T_2 \\ (M_z - M_0)/T_1 \end{pmatrix} \tag{A1}$$

where $\gamma$ is the gyromagnetic constant (42.58 MHz/T for water proton). $M_0$ is the net magnetization of proton spins at equilibrium. T1 is the longitudinal relaxation time. T2 is the transversal relaxation time. $\vec{\Delta B}$ is the local field inhomogeneity. Without applying gradients, the above equation can be further simplified as:

$$\vec{M}(t + \Delta t) = R_{\text{off}} R_{\text{relax}} R_{\text{RF}} \vec{M}(t) \tag{A2}$$

where $R_{\text{off}} = R_{\text{rot, z}}(\theta_i)$ is a matrix describing the rotation around z axis with a counterclockwise angle of $\theta_i = \gamma \Delta B \Delta t$. Rotations around x, y, and z axis with a counterclockwise angle of $\theta$ are defined by the following equations, respectively:

$$R_{\text{rot, x}}(\theta) = \begin{pmatrix} 1 & 0 & 0 \\ 0 & \cos\theta & -\sin\theta \\ 0 & \sin\theta & \cos\theta \end{pmatrix}$$

$$R_{\text{rot, y}}(\theta) = \begin{pmatrix} \cos\theta & 0 & \sin\theta \\ 0 & 1 & 0 \\ -\sin\theta & 0 & \cos\theta \end{pmatrix}$$

$$R_{\text{rot, z}}(\theta) = \begin{pmatrix} \cos\theta & -\sin\theta & 0 \\ \sin\theta & \cos\theta & 0 \\ 0 & 0 & 1 \end{pmatrix}$$

$R_\text{relax}$ is the relaxation matrix defined by:

$$R_\text{relax} = \begin{pmatrix} e^{-\Delta t/T_2} & 0 & 0 \\ 0 & e^{-\Delta t/T_2} & 0 \\ 0 & 0 & e^{-\Delta t/T_1} \end{pmatrix}$$

$R_\text{RF} = R_{\text{rot},z}(\phi) R_{\text{rot},y}(\beta) R_{\text{rot},x}(\alpha') R_{\text{rot},y}(-\beta) R_{\text{rot},z}(-\phi)$, where $\phi$ is the RF phase; $R_{\text{rot},y}(\beta)$ defines a rotation around y axis with a counterclockwise angle of $\beta$; $R_{\text{rot},x}(\alpha)$ is the rotation around x axis with an angle of $\alpha'$; $\beta$ is the angle between x axis and the effective magnetic field [8]; $\alpha'$ is the flip angle around the effective field. $\beta$ and $\alpha'$ are calculated as [9]:

$$\beta = \tan^{-1}(\frac{\tau \Delta \omega}{\alpha})$$

$$\alpha' = -\Delta t \sqrt{(\gamma \Delta B)^2 + (\alpha/\tau)^2}$$

where $\tau$ is the RF pulse duration. When off-resonance or $\tau$ is small, $\beta$ will be close to 0, and $R_\text{RF}$ can be simplified to $R_\text{RF} = R_{\text{rot},z}(\phi) R_{\text{rot},x}(-\alpha) R_{\text{rot},z}(-\phi)$ (the minus sign before $\alpha$ is because $\alpha$ is defined clockwise).

Figure legends

Fig. 1. Illustration of a 1D MRF ZOOM. Starting from an initial place as marked by C (ZOOM center), MRF ZOOM compares the left (L) and right (R) end of the current ZOOM FOV with the center and move the ZOOM to be centered at the maximum (in terms of maximal CC) of the 3 locations until C remains the maximum. In this figure, MRF ZOOM starts from the location of ZOOM1 and moves to ZOOM2, similar to a hill-climbing process as indicated by the arrow.

Fig. 2. Illustration of a 2D MRF ZOOM. With an initial searching resolution, MRF ZOOM starts dictionary generation and matching in the center of the original searching range. It will then move the ZOOM to be centered at the maximum of the boundary points (the green circles and purple squares) and repeat the above matching process until reaching a place (the thick black big square in the figure) where the ZOOM center has a CC greater than the 4 boundary positions (marked by green circles). Then a finer resolution is used to repeat a new MRF ZOOM process until the specified resolution is reached. The white arrow indicates the ZOOM moving direction; the green circles indicate the boundary points that are closest to the center; purple squares indicate the corner points. The dashed big square indicates an intermediate location of ZOOM.

Fig. 3. CC as a function of df, T1, and T2. The MR fingerprints were generated using T1/T2/df=1400/500/100 (msec/msec/df). A) CC vs df curves of all evaluated T1/T2 values, B) CC vs (T1, T2) map at df=100 Hz.

Fig. 4. Computation time of different MRF parameter quantification algorithms. ORIGINAL means the original brute-force algorithm, MZfulldict means MRF ZOOM with a full pre-defined dictionary, MZnodict means MRF ZOOM without using any predefined dictionary, MZdfdict means MRF ZOOM with a predefined dictionary generated with a fixed T1/T2 but with different df values. Dictionary generation time for the prior dictionary (ORIGINAL, MZfulldict, MZdfdict) was not counted in the graph. Error bars mean standard deviation.

Fig. 5. MRF ZOOM results for one brain slice. The left column is the gold standard of T1/T2/df maps. The second column is the MRF ZOOM identified T1/T2/df maps. And the right column is the difference between MRF ZOOM results and the reference.

Fig. 6. CC as a function of df, T1, and T2 when noise presents in MR fingerprints. The MR fingerprints were generated using T1/T2/df=1400/500/100 (msec/msec/df). The noise was set to reduce the auto-CC (between the same MR fingerprints with and without noise) to be 0.4. A) CC vs df curves of all evaluated T1/T2 values, B) CC vs (T1, T2) map at df=100 Hz.

Fig. 7. MRF parameter quantification errors at different noise levels. A) quantification errors of T1/T2/df by different algorithms; B) quantification errors by MRF-ZOOM with and without 3 points Gaussian smoothing.